\documentclass[final,5p,times,twocolumn]{elsarticle}

\usepackage{amssymb}

\usepackage{lineno}

\usepackage{lmodern}
\usepackage{graphicx}
\usepackage{xcolor}
\usepackage{amsmath}
\usepackage{array}

\journal{arxiv}

\begin{document}

\begin{frontmatter}



\title{Absolute quantification of real-time PCR data with stage signal difference analysis}


\author[ciac]{Chuanbo Liu}
\author[ciac,stonybrook]{Jin Wang\corref{cor1}}
\cortext[cor1]{To whom correspondence should be addressed. Email: jin.wang.1@stonybrook.edu}
\address[ciac]{State Key Laboratory of Electroanalytical Chemistry, Changchun Institute of Applied Chemistry, Chinese Academy of Sciences, Jilin, People's Republic of China}
\address[stonybrook]{Department of Chemistry, Physics and Applied Mathematics, State University of New York at Stony Brook, Stony Brook, USA}

\begin{abstract}
  Real-time PCR,
  or Real-time Quantitative PCR (qPCR) is an effective approach to quantify nucleic acid samples.
  Given the complicated reaction system along with thermal cycles,
  there has been long-term confusion on accurately calculating the initial nucleic acid amounts from the fluorescence signals.
  Although many improved algorithms had been proposed, 
  the classical threshold method is still the primary choice in the routine application. 
  In this study,
  we will first illustrate the origin of the linear relationship between the threshold value and logarithm of the initial nucleic acid amount by reconstructing the PCR reaction process with stochastic simulations. 
  We then develop a new method for the absolute quantification of nucleic acid samples with qPCR.
  By monitoring the fluorescence signal changes in every stage of the thermal cycle, 
  we are able to calculate a representation of the step-wise efficiency change.
  This is the first work calculated PCR efficiency change directly from the fluorescence signal, 
  without fitting or sophisticated analysis. 
  Our results revealed that the efficiency change during the PCR process is complicated and can not be modeled simply by monotone function model. 
  Based on the calculated efficiency, 
  we illustrate a new absolute qPCR analysis method for accurately determining nucleic acid amount.
  The efficiency problem is completely avoided in this new method.
\end{abstract}


\end{frontmatter}


\section{Introduction}
\label{S:1}

Ever since the invention of Polymerase Chain Reaction (PCR) \cite{Mullis1986},
and subsequently the development of quantitative PCR (qPCR) \cite{Holland1991, Higuchi1993},
it has been applied to substantial areas involving the amplifications of DNA/RNA sequence and
the characterization of amount of specific molecules that contained in the sample solution.
The power of qPCR lies in the exponential-like high products/sample ratio,
or in other words the ability for amplification of sample signal in an exponential manner.
With the mechanism of sequence complementarity of DNA molecules, 
the signal can be amplified with high specificity. 
These features give qPCR the ability to accurately detect low concentration sample with great specificity when combining with fluorescence technique,
such as passive fluorescence dyes,
 or fluorescent reporter probe or short F\"{o}rster resonance energy transfer probes \cite{Kutyavin2013}.
As a direct result,
the step-wise fluorescence signal is expected to be linearly proportion to the amount of the double-strain DNA (dsDNA) presented under the same thermodynamic conditions.

The currently widely employed qPCR data analysis algorithms for absolute and relative quantifications relay on several assumptions which have been questioned during the last decade.
The first and most strong assumption assumes an exponential increase of the products with constant amplification efficiency.
The overall exponential increment is clearly not possible for a finite closed system.
In fact the decreasing of enzyme activity, the exhausting of reaction substrates,
the competition of the binding sites for high concentration reagents all lead to the cease of amplification,
and hence the appearance of the plateau in large cycle number.
Therefore, the assumption was restated in a weaker version that only assumes exponential increase with the constant efficiency in the beginning of the reaction cycles.
By carefully choosing a threshold value $C_T$ for each amplification profile,
linear relationship can be established between $C_T$ and the logarithm of the initial amount of the target sample.
Therefore a standard curve can be established by linear regressive analysis.
Averaged efficiency can also be obtained from the slope of the standard curve. 
These together formed the theoretical basis for the quantification algorithms used in many commercial software pre-installed in qPCR instruments.

However, further studies showed the potential deviations of this method.
The problems mainly come from two aspects, 
the baseline and the efficiency calculated from fluorescence signals.
A uniform, minimal fluctuated baseline is crucial for choosing proper $C_T$.
Reference dye such like Rox is usually used to normalize fluorescence signal and to correct well-to-well optical variations.
Instructions should also be calibrated and maintained carefully to achieve minimized fluorescence signal fluctuations.
Moreover, in order to satisfy the constant exponential increment assumption, 
the choice of $C_T$ is usually made at a small region just above the baseline,
such as 10 times baseline fluctuations from the practical advices. 
But manual adjustments were often made during data analysis, 
especially for large fluctuated signals or high initial concentrations,  
where the first cycle would produce a signal far beyond the threshold region for standard samples.
Even with these precautions, 
the estimated baseline still gives high potential errors to the qPCR results.
In fact,
the baseline is estimated in the early cycle number and extended to the end,
which is actually the averaged baseline without step-wise fluctuations.
Therefore there is no surprise to see the estimation of baseline has large impacts on the observed efficiency due to exponential-like behavior of the reaction cycle \cite{Ruijter2009}. 
A recent discovery suggested that the fluorescence signals of baseline had large non-random fluctuations,
highly related to the sample natures.
In fact,
the scale of the fluctuations was observed to increase with increasing cycle number \cite{Spiess2016}.

The baseline problem is not severe to the linearity of a standard curve.  
Methods have been developed to calculate the fractional $C_T$ insensitive to baseline selection using the first or the second derivatives to obtain an invariant value \cite{Luu-The2005}.
However, the efficiency difference can give arise to large deviation for the calculation of initial concentration even with a perfect linear standard curve. 
In the classical threshold methods that applied widely, 
amplification efficiency is assumed not changing before the threshold level. 
Therefore, the initial sample amount can be calculated from the standard curve and the measured threshold value.
However, simulation and experiment results indicate the efficiency change is as complex as the amplification fluorescence signals, 
and may have already decreased 10--25\% below maximum value at the time when amplification signals are distinguishable from the background \cite{Rutledge2008,Gevertz2005}.
Besides, it is commonly known the amplification efficiency varies with product length and sequence. 
For different amplification target sequences or the same target sequence with different contexts,
efficiencies can be significantly different from each sample, leading to large estimation errors due to the exponential-like behavior of the thermal reaction cycle.
So the problem is,
even if we had a perfect linear relationship, 
the initial sample amount was still unreachable since we did not have any knowledge of the efficiency change. 
Therefore the classical threshold methods is in principle failed in calculating the sample amount with different initial efficiency.
Although sequence difference related efficiency change can be circumvented with the relative quantification using efficiency corrections \cite{Pfaffl2001},
the calculation was still relied on averaged efficiency. 
Also, even though the relative method corrects efficiency divergence of the samples to certain extent but the method is highly dependent on the quantity of the reference gene.
Typically, non-regulated genes or housekeeping genes are used for references.
However, studies had shown housekeeping genes are also being regulated and may vary under experimental conditions \cite{Thellin1999}.

Mechanistic models with dynamic efficiency were also developed to fit individual qPCR fluorescence signal curve globally.
Sigmoid and other S-shaped models were examined to fit the qPCR curve \cite{Rutledge2004,Guescini2008}.
A so-called ``Full Process Kinetics-PCR'' (FPK-PCR) combined the Sigmoid fitting and the second derivative method together to plot the standard curve for quantification \cite{Lievens2011}.
The results did show insensitivity to the efficiency of different samples or the presentation of inhibitors,
but the collective studies indicate that the $C_T$ method is more precise for absolute quantification than 11 mechanistic models \cite{Goll2006}.
More importantly, fitting of individual amplification curve may introduce random error and increase ambiguity. 
Therefore it should not be used to replace the classical methods \cite{Bar2011}.

In this study, 
by using a stochastic Langevin model,
we were able to show the robustness of threshold methods in the construction of linear relationship between the threshold values and the logarithm of the initial concentration. 
The origin of the log-linear behavior was attributed to the invariant amplification kinetic constants of PCR reaction cycles.
Threshold log-linear relationship should hold true as long as the thermodynamic factors were not changed. 
Moreover, 
a new method employed the stage difference signal analysis (SSDA) of the thermal reaction was developed to determine the step-wise efficiency change along the reaction process.
Further, 
a new quantification method was also developed using the SSDA method to calculate the accurate initial nucleic acid amount.
The SSDA method recorded the fluorescence signal at the end of each stage of qPCR thermal cycle in contrary to the traditional qPCR which only recorded the fluorescence signal at the end stage of a complete single thermal cycle.
Differences of fluorescence signals of the thermal cycles gave complete information about the whole amplification process.
The baseline was handled naturally and in a step-wise fashion along the whole thermal reaction cycles, 
which resulted in accurate determination of the amplification signal as compared with the averaged baseline estimation procedure. 
Therefore, the efficiency can be calculated independent of the background signals. 
Based on the calculated efficiency change, 
absolute or relative quantification algorithms can be constructed to get precise results of the initial nucleic acid amount in various samples. 
Here we used the calculated efficiencies to obtain the cumulative amplification fold (CAF) as a function of the qPCR cycle number.
Our results showed perfect linear relationship between CAFs and initial sample concentrations in a wide initial concentration range. 
Therefore, the CAFs were proposed to be used as the characteristic values for standard curve plot. 
By using CAFs instead of the fluorescence thresholds that corresponds to total dsDNA amount presented in the measurement wells, 
amplification efficiencies of different samples with distinct sequences can be handled in a step-wise procedure. 
Efficiency problem was solved in principle.
Besides, manual manipulation of the threshold and the baseline subtraction was completely avoided. 

The SSDA methods we developed in this study can be applied readily in most commercial equipments without additional devices. 
The data analysis is also straightforward and high-throughput program can be easily designed. 
Moreover, the method can be extended to determine the relative gene expression level with an arbitrary gene, 
or with a standard sample synthesized \emph{in vitro} which is not possible using threshold algorithms.

\section{MATERIALS AND METHODS}

\subsection{Sample preparation}

Plasmid pEA206 with high replica region and an N-terminal phyB gene were synthesized and transfected into DH5$\alpha$ competent cell for plasmids amplification. 
pEA206 sequence was indicated in the Supplementary Materials. 
Plasmids were then extracted from the E.coli cells and purified for real-time PCR assay using GeneJET Plasmid Miniprep Kit (Thermo Scientific).
All samples were stored in 4 $^{\circ}$C for no longer than a week before experiments. 
Sample concentrations were determined with NanoDrop microvolume UV-Vis spectrophotometer (Thermo Scientific).

\subsection{Real-time PCR assay}

PCR primers were designed targeting the N-terminal phyB gene in pEA206 plasmid. 
Primer sequences were as follows: 
pEA206 forward: 5'-GCGATTTCTCAGTTACAGGCTCTTC-3';
pEA206 reverse: 5'-TACTATCATTCGGACACGGTTCTGC-3' (255 bp product).
Primers were synthesized and purified to HPLC grade from ShengGong, Shanghai. 
Samples and the serial dilution data series were amplified in 96-well plates in an Applied BioSystems StepOne Plus instruments. 
The reaction system were 20$\mu L$, 
and SYBR green MasterMix from TAKARA were employed in all the experiments. 
Thermal cycle protocol was identical for all experiments except temperature variable experiments: 
10 min 94 $^{\circ}$C, 
followed by 45 $\times$ (15 s 94 $^{\circ}$C, 10 s 60 $^{\circ}$C, 1 min 60 $^{\circ}$C, and 10 s 60 $^{\circ}$C).
For temperature variable experiments,
the annealing condition was 58$^{\circ}$C for 20 s and extension stage condition was set at 68 $^{\circ}$C for 1 min. 

\subsection*{PCR efficiency calculation}

Fluorescence signal of PCR data were recorded at the denaturation, 
end of annealing and end of extension stages, 
termed F1, F2 and F3, respectively.
For our qPCR instructions,
10 s was the minimum time interval for data recording. 
Therefore, two additional thermal steps that last 10 s were added before and after the 1 min 60 $^{\circ}$C for fluorescence signal recording.
The efficiency was defined with the step-wise amplification fold before and after extension in each reaction step.
Therefore, efficiency was calculated by the following equation.
\begin{equation}\label{eq:efficiency-calculation}
 E^{(n)} = (F3^{(n)} - F1^{(n)}) / (F2^{(n)} - F1^{(n)}) -1
\end{equation}
where $n$ indicated the cycle number.
The cumulative amplification fold (CAF) which was a function of thermal cycle number $m$ was calculated as
\begin{equation}\label{CAF-definition}
CAF(m) = \prod_{n=0}^{m} (1 + E^{(n)})
\end{equation}

\subsection{Absolute analysis algorithm}

The general formula for calculation of PCR product at certain cycle number was constructed by considering step-wise efficiency change. 
\begin{equation}\label{eq:product_general}
 N_{C_T} = N_0 \cdot \prod_{n=0}^{C_T} (1 + E^{(n)})=N_0 \cdot CAF(C_T)
\end{equation}
where $N_0$ was the initial value.
By only assuming the linear correlation of SYBR green fluorescence signal and the dsDNA presented in system, 
a linear relationship can be constructed between $log(N_0)$ and CAF as defined in Eq. \ref{CAF-definition} as followed.
For some particular threshold $F_{th}$ on the baseline subtracted curve $\Delta F = F3 - F1$,
the amplified products were the same for all the samples being tested. 
So a constant value was assigned to each amplification profile. 
\begin{equation}\label{eq:amplification_to_constant}
  N_0 \cdot CAF(C_T) = const
\end{equation}
By taking logarithm of both sides, 
a linear relationship emerged.
\begin{equation}\label{eq:absolute_std_curve}
 log(CAF(C_T)) = log(const) - log(N_0)
\end{equation}
Therefore a standard curve can be plotted with respect to various initial concentrations. 
Since efficiency change was given by the SSDA method, Eq. \ref{eq:absolute_std_curve} was not related to specific sample natures. 
So nucleic acid samples with distinct sequences can be quantified regardless of the different amplification efficiencies. 

Since the exponential amplification manner of qPCR assays, 
errors were also accumulated throughout the thermo cycle. 
So $C_T$ values determined in the early stage of amplification helped in  minimizing the accumulated errors.

\subsection{Stochastic simulations}

Stochastic simulations were performed with a simple recursive reversible two-state kinetic model with invariant reaction constants. 
For each thermal reaction cycle, 
the reversible reaction can be written by the following element equations:
\begin{align}\label{sim-eq}
    P + T \to 2T \\
    2T \to P + T
\end{align}
where P stands for primer pairs,
and T stands for the product template. 
The efficiency can be calculated by the division of product amount before and after each thermal cycle.
The result showed dependence on both the forward reaction rate $k_1$ and backward reaction rate $k_2$.
\begin{equation}
 T^{(n + 1)} / T^{(n)} = 1 + k_1 \cdot P^{(n)} - k_2 \cdot T^{(n)} 
\end{equation}
By the definition of efficiency, it can be written explicitly by the following equation.
\begin{equation}
  T^{(n + 1)} / T^{(n)} = 1 + E^{(n)}
\end{equation}
and the efficiency equalled to,
\begin{equation}\label{sim-efficiency}
 E^{(n)} =  k_1 \cdot P^{(n)} - k_2 \cdot T^{(n)} 
\end{equation}
Randomness were account for the fluorescence signal fluctuations,
the step-wise product fluctuations,
and the measurement bias.
For simplicity, 
we employed an Langevin equation instead of Gillespie algorithm to account for the noise explicitly. 
The Langevin equations were obtained by adding a step-wise Guassian noise item to the deterministic recursive kinetic model. 
\begin{equation}\label{eq: Langevin}
  T^{(n+1)} = T^{(n)} + \sigma \cdot noise
\end{equation}
where $\sigma$ is the scale factor for the fluctuation or noise strength.
The primer pairs after the thermal cycle taking into account of the efficiency fluctuation is related to the initial amount of primer pairs and the sample amount by the following equation.
\begin{equation}
 P^{(n)} = P^{(0)} + T^{(0)} - T^{(n)}  
\end{equation}
where $P^{(0)}$ and $T^{(0)}$ stand for the initial amount of primer pairs and dsDNA template, respectively.
The fluorescence signal generated by this recursive reversible two-state model can be calculated by the following equation.
\begin{equation}
 F^{(n)} = F_0 + \Delta F \cdot T^{(n)} + \gamma \cdot noise 
\end{equation}
where $F_0$ represents the background fluorescence signal,
$\Delta F$ represents the fluorescence signal increment per template product,
and $T$ represents the amount of dsDNA template.
$\gamma$ represents the fluorescence fluctuation level factor.
Kinetic parameters and the fluctuation or noise strength were tuned according to previous simulation results \cite{Cobbs2012}.

\section{RESULTS}

\subsection{Original and robustness of log-linear relationship}

qPCR is a complicated system involving the correlated interactions of the template,
the primer pairs,
the enzyme and buffer system that contains essential ions to ensure reactions to occur within each thermal cycle.
Relatively few mathematics model had been proposed for the analysis of qPCR system,
partly because the system is quite complicated, 
another reason is the simulation often has limit effects on the precision of the calculated results.
Both deterministic and stochastic method has been constructed,
using the law of mass action \cite{Stone2006}, hidden markov chain model or Bayesian method \cite{Lalam2008,Sivaganesan2008}.
Discrete cycle numbers using recursive deterministic reaction model were also developed to simulate and fit the qPCR data globally \cite{Carr2012}.
Sequence variations were analyzed with a thermodynamic deterministic model \cite{Marimuthu2014}.
These results yield insights into the mechanism of qPCR reactions, 
but none of them give the clues about why log-linear relationship emerges. 

\begin{figure}[t]
    \begin{center}
    \includegraphics[width=9cm]{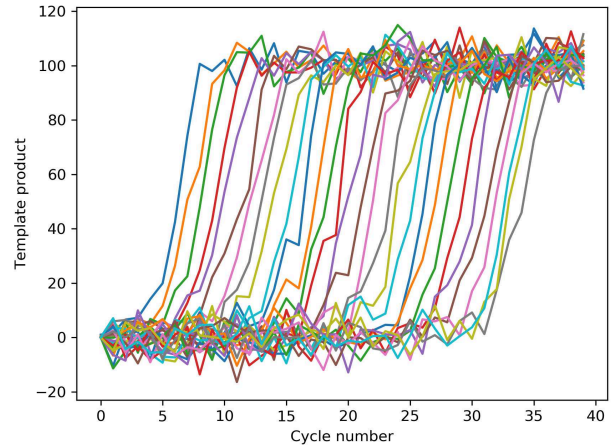}
    \end{center}
    \caption{
      \textbf{Stochastic simulation of qPCR data with recursive reversible two-state model.}
     30 data trajectories were generated with the same kinetic rate constant. 
    Fluorescence fluctuations as well measurement bias and efficiency deviations were considered in the generated trajectories.
    Initial template copy number was set to be 1 with total primer pairs copy number equalled to 100.
    Signal fluctuation level was set to be $\gamma$ = 5,
    efficiency fluctuation level which is highly sensitive due to the exponential accumulative manner was set to be $\sigma = 10^{-8}$.}
    \label{fig:simulation_trajectories}
\end{figure}

\begin{figure}[t]
    \begin{center}
    \includegraphics[width=9cm]{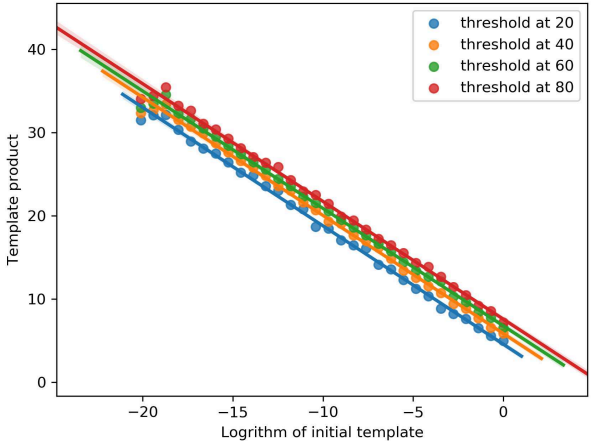}
    \end{center}
    \caption{
      \textbf{Threshold value plot with logarithm of initial template amount.}
      Threshold values were calculated with interpolate of the simulated curve with threshold level set to be 20, 40, 60, and 80, respectively. 
      Linear regression with deviations were plotted for each threshold level. 
    }
    \label{fig:simulation_threshold}
\end{figure}

In the current study we adopted a stochastic method of a recursive reversible two-state model including a noise term accounting for the fluctuating effects, 
as shown in Eq. \ref{eq: Langevin}.
The main advantage of using stochastic method is the ability for addressing the role of noise in the qPCR reactions and data recording.
The study here obviously does not have the exactly right mathematic model for qPCR simulation,
but we employed the model here to show that the origin of log-linear relationship is the repetitive reaction behavior of qPCR, 
not from the conventional thought of constant exponential product increment.
The simulation results that generated from Eq. \ref{sim-eq} with both the fluorescence fluctuation and efficiency deviation were shown in Fig. \ref{fig:simulation_trajectories},
and threshold values calculated with the interpolations of the simulated trajectories were shown in Fig. \ref{fig:simulation_threshold}. 
The results clearly illustrate the origin of success of the classical threshold method. 
Even with large fluorescence and efficiency fluctuations, 
a good performed standard curve can still be obtained. 
Given the large range of the initial template copy numbers,
and the insensitive against the location of threshold level, 
the simulation results here explain the robustness of the classical threshold method.
However, 
it can be seen that the linear relationship failed at very low initial copy numbers due to efficiency fluctuations.
The reason can be well understood since the cumulative nature of PCR reaction cycle will amplify the signals as well as the efficiency errors.

It had to be noted that,
the model we employed here was clearly not the optimized one for simulating qPCR reactions.
However, we still got a well defined standard curve which showed log-linear relationship with the initial conditions.
The only invariant parameters were the kinetic rate constants.
If the rate constants were changed during the simulation, 
log-linear relationship was broken.
Therefore, 
the origin of the log-linear relationship can be well explained by the repetitive reaction cycles following exactly the same yet complicated reaction kinetics mechanism.
No wonder we can always achieve the good performed linear relationship despite the various algorithms that were used to calculate the threshold values.
Hence the discussions here illustrate the mystery of the successes of various threshold methods currently widely used by commercial instructions and softwares.

\subsection{Obtain step-wise efficiency change with stage signal difference method}

The traditional qPCR data acquired method focused on the fluorescence signals at the end of each thermal cycle.
Given the data recording process that usually takes a time period of 10 seconds, 
the recorded fluorescence signals are actually the averaged signals together with all the constructions from the reaction system. 
While many commercial instructions give the flexible methods that allow recording fluorescence signal at any stage of the thermal reaction cycles, 
the potential of qPCR methods were not fully utilized since many signals were unrecorded. 

\begin{figure}[t]
    \begin{center}
    \includegraphics[width=9cm]{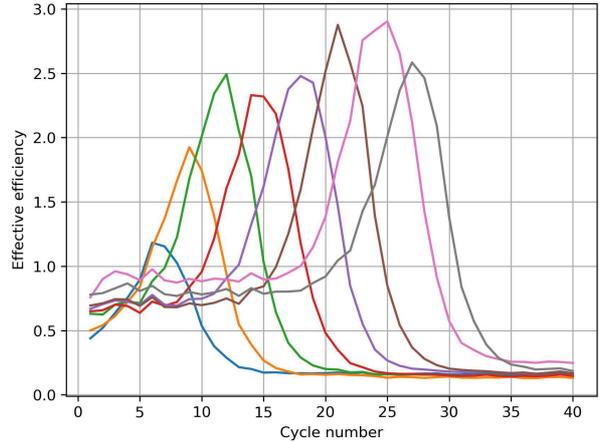} 
    \end{center}
    \caption{
      \textbf{Efficiency calculated from the fluorescence signals before and after extension.}
      The efficiencies were calculated from Eq. \ref{eq:efficiency-calculation}.
      Experimental conditions and procedures were the same as listed in the methods part. 
    }
    \label{fig:efficiency_plot}
\end{figure}

Given these circumstances, 
we demonstrated here that step-wise efficiency can be achieved by recording the fluorescence signals before and after extension stage. 
The fluorescence signals at the denaturation stage were also recorded. 
For a routine PCR reactions, the denaturation stage was usually set at 92-98$^{\circ}$C and it lasted for about 10 minutes. 
At the end of this stage,
dsDNA can be considered completely denatured. 
Therefore the fluorescence signals of the denatured stage can be viewed as a good description of the background signals for each thermal cycle. 
After subtracting the background signals, 
fluorescence signals reflect purely the amount of dsDNA currently presented in the reaction system. 
If the data recording can be performed instantaneously, 
the signals can be recorded at the moment that the temperature is dropped to annealing and extension stage, 
representing the previously exist amount of dsDNA. 
Thus the signal recorded at the end of extension stage can represent the existing amount of dsDNA after extension. 
The division of the signals recorded after extension with the signals recorded just before the annealing gave the amplification efficiency of the current thermal cycle step.
As a conclusion, we estimated the step-wise efficiency according to Eq. \ref{eq:efficiency-calculation}.

As shown in Supplement Figure 1,
fluorescence signal from 5 different reaction stages were recorded to illustrate the fluorescence signal change during the whole PCR reaction cycles. 
Further in Supplement Figure 2, 
the fluorescence signals before and after extension shown different fluorescence signals as expected from the synthesized dsDNA fragments. 
The calculated efficiencies are shown in Fig. \ref{fig:efficiency_plot}. 
It is surprising to see the efficiencies are not at their maximum at the beginning of the reaction cycles but rather, 
have small efficiency values comparable to the plateau phase. 
This result is contrary to all the known simulation results and hypothesis that we used for the calculating of initial sample amount from standard curve methods. 
The amplification efficiency reached its maximum in the middle of the thermal cycle procedure, 
this is in accordance with the observation that fluorescence signal changed most drastically in the so-called log-linear phase of amplification. 
It has to be noted that, 
the efficiency we calculated here might not be the real efficiency for each thermal cycle. 
We have no idea whether the fluorescence signals follow the same linear law in every reaction stage. 
Therefore, in the circumstances that different linear relationships occurred in different PCR stages,
assume $F1^{(n)} = b$, $F2^{(n)} = a' \cdot x + b$, and $F3^{(n)} = a'' \cdot y + b$, 
where $x$ and $y$ are the effective amount of nucleic acid present in the solution, $a'$ and $a''$ are the parameters relate fluorescence signals to dsDNA amount in solution. 
Then the calculated efficiency, $E_c$, equals to $\frac{a''}{a'} \cdot \frac{y}{x} - 1$, the real efficiency, $E_r$, equals to $\frac{y}{x} - 1$. 
The relationship between $E_c$ and $E_r$ are $E_c = \frac{a''}{a'} \cdot (E_r + 1) - 1$.
So the efficiency calculated here were linear related to the real efficiency. 
But since the linear relationship of calculated efficiency and real efficiency is the same in one particular reaction for all the tested samples, 
an absolute quantification method can still be established.

\begin{figure}[t]
    \begin{center}
    \includegraphics[width=9cm]{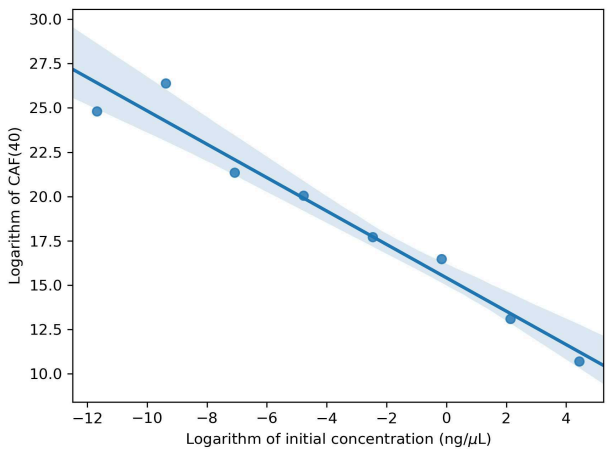} 
    \end{center}
    \caption{
      \textbf{Plot and linear regression of total amplification fold with sample initial concentrations.}
      The total amplification fold were calculated with Eq. \ref{CAF-definition}. 
      The samples were range from 84.6ng/$\mu$L to 84.6 $\times\ 10^{-7}$ ng/$\mu$L. 
      Each sample was prepared by serial diluting 10 folds from the previous sample. 
      The confidence interval calculated using a bootstrap for regression estimation was drawn with translucent blue bands around the regression line for better visualization. 
    }
    \label{fig:amp_with_inital}
\end{figure}

To verify if the efficiencies calculated really represented the efficiencies for each sample, 
The CAF was calculated with Eq. \ref{CAF-definition}. 
The resulted CAF was plotted with the initial concentrations as shown in Fig. \ref{fig:amp_with_inital}. 
The data points were fitted with linear regression.
As we have discussed previously, 
the calculated efficiencies were linear related to the real efficiencies, 
so the fitting slope was not equal to 1 like Eq. \ref{eq:absolute_std_curve}.
The results clearly indicate the linear relationship of calculated CAF with the initial concentrations, 
confirmed the amplification fold nature of the cumulative product of calculated efficiencies. 
While for the small concentration samples, 
the divergence of linear relationship can be explained by the accumulation of errors for very diluted samples.

In summary, 
the signal difference before and after extension stage provides a representation of the step-wise efficiency change along the thermal reaction cycle. 
Noted that, to our best knowledge, this is the first reported study on directly calculating the PCR efficiency change from fluorescence signals. 
The results shown here seem in contradiction to almost all the other simulation results, 
indicating the very complex nature of the PCR reactions system.

\subsection{Absolute quantification}

As shown in the stochastic simulations, 
a standard curve is very robust for a fixed reaction rate system with repetitive recursive cycles. 
Therefore, a standard curve was still employed to minimize the external errors and increase the robustness of the analysis algorithm.
The fundamental difficulty of classical absolute quantification methods are efficiency difference between standard and test samples. 
Although a good looking standard curve can always be obtained, 
calculated initial amount of test samples are not reliable. 
By handling efficiency change step-wise with SSDA methods and following the simple assumption that the fluorescence signal intensities were proportional to PCR products presented in solution for the same reaction system, 
cumulative amplification fold (CAF) can be calculated. 
Obviously the CAF curves are different for standard and test samples, 
however the relationship between CAF and $N_0$ hold true despite the sample difference as shown in Eq. \ref{eq:product_general}.
Following this argument, the standard curve was calculated as illustrated in Eq. \ref{eq:absolute_std_curve}. 
In this way, 
the efficiency problem was handled naturally with minimum assumptions. 

In order to illustrate the influence of threshold positions, 
5 thresholds were selected spanning the whole amplification process with same intensity distances. 
Crossing points were obtained by locating the first intersect point between threshold line and fluorescence difference $F3 - F1$ as defined in Eq. \ref{eq:efficiency-calculation}. 
Fractional cycle numbers that obtained from the x-coordinates of crossing points were assigned to $C_T$ values. 
At last, CAF values were interpolated from the $C_T$ values. 
Standard curves corresponding to different thresholds were plotted between logarithm of CAF values and logarithm of the initial concentration for visualization. 

\begin{figure}[t]
    \begin{center}
    \includegraphics[width=9cm]{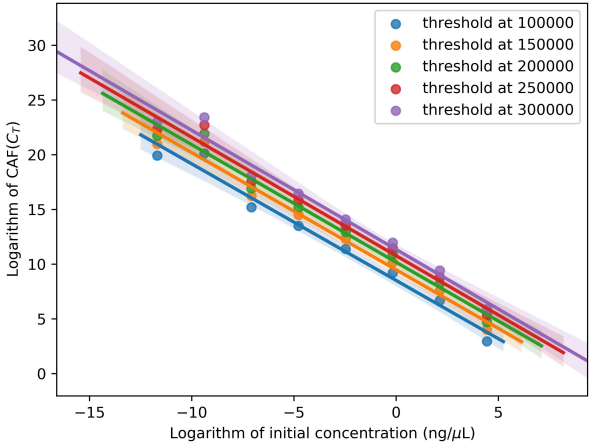} 
    \end{center}
    \caption{
      \textbf{Plot of standard curves for different threshold levels.}
      The cumulative summation of efficiencies were calculated with Eq. \ref{CAF-definition}. 
      $C_T$ were determined by setting thresholds at 100000, 150000, 200000, 250000 and 300000. 
      The values of the standard points were obtained by interpolation between the CAF values around the cross points. 
      Linear regression were performed in the general two parameters form of Eq. \ref{eq:absolute_std_curve} and labeled with different colors.
      Standard deviation were calculated and plotted with light blue. 
    }
    \label{fig:general_std_curve}
   \end{figure}

As shown in Fig. \ref{fig:general_std_curve}, 
well defined linear standard curves can be obtained. 
The efficiencies change were handled naturally in the cumulative summation according to the efficiency profiles. 
Since the efficiency profiles were obtained step-wise in our SSDA method, 
efficiency difference in samples had no impacts on the calculated initial nucleic acid copy numbers.
The initial concentration of standard samples were 84.6 ng/$\mu$L. 
And serial dilution were done with each standard samples diluted by factor of 0.1. 

\begin{table}[b]
    \begin{center}
        \caption{Linear regression results of calculated standard curves from SSDA methods.Threshold values were determined from fluorescence signals after extension.}
        \label{table:std_fit_para}%
        \begin{tabular*}{\columnwidth}{@{}ccccc@{}}
            \hline
            Threshold & Slope & Intercept & R-square \\
            \hline
            100000 & -1.066 $\pm$ 0.065 & 3.694 $\pm$ 0.181 & 0.975 \\
            \hline
            150000 & -1.072 $\pm$ 0.042 & 4.117 $\pm$ 0.178 & 0.975 \\
            \hline
            200000 & -1.075 $\pm$ 0.064 & 4.417 $\pm$ 0.179 & 0.975 \\
            \hline
            250000 & -1.083 $\pm$ 0.066 & 4.675 $\pm$ 0.184 & 0.974 \\
            \hline
            300000 & -1.089 $\pm$ 0.068 & 4.925 $\pm$ 0.191 & 0.973 \\
            \hline
        \end{tabular*}%
    \end{center}
\end{table}

The slope, intercept and R-square that calculated from linear regression were shown in Table \ref{table:std_fit_para}. 
The results showed the robustness of the new methods as expected from the threshold standard curve methods. 
Notice that all the standard curves had similar slope and intercepts were related to the position of the thresholds. 
The R-squares were not perfect due to the large deviations in the diluted samples as a result of the accumulated errors in CAF calculation. 
More sophisticated arrangement of experiments may give better results. 
Besides, it has to be noted that the slopes calculated were related to the coordinates so the calculated slopes had nothing to do with Eq. \ref{eq:absolute_std_curve}.

\subsection{Influence of temperature difference in annealing and extension stages}

In practical application of qPCR, 
especially for long PCR products, 
an extension stage with higher temperature is often useful for achieving optimized enzyme activity. 
Therefore, discussion about the temperature influence on SSDA methods is valuable for temperature invariant method.  

\begin{figure}[t]
    \begin{center}
    \includegraphics[width=9cm]{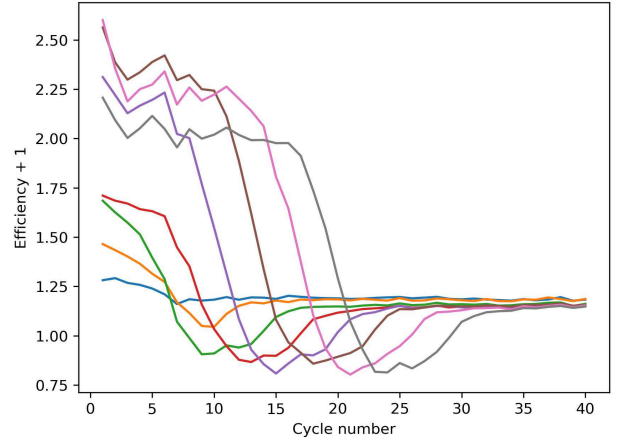}
    \end{center}
    \caption{
      \textbf{Temperature difference effects on efficiency determination.}
      The fluorescence signals were recorded during annealing and extension stages.
      Annealing and extension were performed at 58$^{\circ}$C and 68$^{\circ}$C, respectively.
      Efficiency were calculated according to Eq. \ref{eq:efficiency-calculation}.
      Results were moved upward to get rid of the negative values.
    }
    \label{fig:temp_diff_efficiency}
\end{figure}

An example experiment was performed to show the impacts of temperature variations on efficiency determination. 
The `efficiency' calculated from Eq. \ref{eq:efficiency-calculation} was shown in Fig. \ref{fig:temp_diff_efficiency}. 
It can be noticed that the efficiency dropped below zero at the middle of the thermal cycles indicating the difference of temperature between annealing and extension showed profound impacts on the calculated results. 
Further analysis showed the minimum of the efficiency was right at the position of the maximum of fluorescence signal differentiation as shown in Supplement Figure 3.
The explanation of this result can be attributed to the competition between PCR produced dsDNA and the thermal denatured dsDNA in the system. 
The PCR produced dsDNA kept dropping as the enzyme activity, substrates concentration were dropping, 
while the thermal denaturation effects was not changed but consistently denature a constant portion of dsDNA in current thermal cycle. 
Therefore, a turning point was expected at the maximum amplification rate cycle. 
This explained the temperature effects on efficiency determination. 
The net result was, 
a constant decrease factor was added to the calculated efficiency profile which was the same for all the standard and test samples. 
Therefore the stage signal difference analysis results no long lead to the representation of the efficiencies.
However, quantification methods can still be established since the denaturation portions are the same for all the samples. 
Therefore the calculated initial copy numbers are not changed with temperature difference stages. 
The only impact of temperature effects was introducing external factors that making the efficiency representation failed.

\section{DISCUSSION}
qPCR is a widely applied method for determination of low concentration nucleic acid samples with extremely high specificity. 
However, the method for analyzing acquired data suffered a decade debates about the fundamental theory basis and precisions in the calculation results. 
In this paper, 
the origin of the log-linear relationship in threshold methods was illustrated with stochastic simulations. 
The fundamental reason for a linear relationship was uncovered to be the same kinetic feature for a repetitive reaction cycle. 
Therefore the standard curve obtained from various threshold levels all exhibited the same features,
such as the same slope but with different intercept values. 
This explained the validations of various threshold value calculation methods, 
and proved the robustness of the threshold methods. 
Furthermore, 
we developed a new method called stage signal difference analysis (SSDA) method for obtaining step-wise efficiency change along the thermal cycles. 
This new method takes advantage of the full information that a thermal cycle can provide to get a full description of the product dsDNA change. 
Step-wise efficiency can be calculated very easily and a well defined standard curve can be used to give the absolute quantification of nucleic acid samples. 
Efficiency and baseline problem that bothered the classical threshold methods were completely avoided, 
therefore precises results can be achieved across difference samples with different contexts and sequences. 

The method we developed here can also be used to construct a relative measurement algorithm. 
For example, 
the relative ratio between standard and test sample can be calculated based on the CAF of two different genes. 
\begin{equation}\label{eq:relative_quantification}
    ratio = \frac{CAF(gene1, C_T)}{CAF(gene2, C_T)}
\end{equation}
where $gene1$ and $gene2$ represent the target gene and reference gene, respectively. 

Further developments and validations on the new SSDA method may be tested on different samples and architecture environments to estimate the results calculated from Eq. \ref{eq:relative_quantification}. 
More sophisticated method for absolute and relative quantification are possible based on the SSDA method we developed here.
The results shown in this study were performed with SYBR green I, 
other fluorescence signal monitoring methods will have no difficulty in applying the SSDA methods. 

To our best knowledge, this is the first time that a full information recording qPCR analysis method reported. 
It opens the way for more sophisticated analysis methods developed for exponential-like mechanism reactions, 
such like the rolling circle replication reaction and other signal amplification methods for small amount of sample detection.

\section{CONCLUSION}

As a conclusion, 
a standard curve method based on SSDA method was developed for absolute quantification of nucleic acid samples. 
This method focused on the cumulative summation of efficiency profile rather than the characteristic value $C_T$ calculated from simple threshold procedure. 
The efficiency and baseline problems were handled naturally during the calculation of standard curve, 
so no theoretical deviations existed for calculating different samples that had different amplification efficiencies. 
Therefore, this method was intrinsically insensitive to the sample difference as contrary to the classical threshold method.
The method was simple and easy to perform, 
and even eliminated the use of reference dye as normalization factor. 
High-throughput methods can also be implemented with simple codes.
The robustness of this method was expected to be no less than the classical threshold method. 
This is the first reported qPCR data analysis methods that attempting to utilize all the fluorescence signal produced during the PCR thermal cycles. 
The method introduced here revealed the great potentials hidden in real-time fluorescence recording of complex chemical reactions. 

\section{ACKNOWLEDGEMENTS}

Chuanbo Liu thanks supports by Natural Science Foundation of China, No.32000888,
Jin Wang thanks the supports from grant no. NSF-PHY 76066 and NSF-CHE-1808474,
Ministry of Science and Technology of China, No.2016YFA0203200, 
Projects of Science and Technology Development, Jilin, China, No.20180414005GH,
Projects of Instrument and Equipment Development, Chinese Academy of Sciences, No.Y928041001. 






\begin{thebibliography}{99}

\bibitem{Mullis1986}
Mullis,K., Faloona,F., Scharf,S., Saiki,R., Horn,G. and Erlich,H. (1986)
Specific enzymatic amplification of DNA in vitro: the polymerase chain reaction.
\textit{Cold Spring Harbor Symposia on Quant. Biol.},
\textbf{LI}, 
263--273.

\bibitem{Holland1991}
Holland,P.M., Abramson,R.D., Watson,R. and Gelfand,D.H. (1991)
Detection of specific polymerase chain reaction product by utilizing the 5'$\rightarrow$3' exonuclease activity of \textit{Thermus aquaticus} DNA polymerase.
\textit{Proc. Natl. Acad. Sci. U.S.A.},
\textbf{88}, 
7276--7280.

\bibitem{Higuchi1993}
Higuchi,R., Fockler,C., Dollinger,G., and Watson,R. (1993)
Kinetic PCR Analysis: Real-time Monitoring of DNA Amplification Reactions.
\textit{Nat. Biotechnol.},
\textbf{11},
1026--1030.

\bibitem{Kutyavin2013}
Kutyavin,I.V. (2013)
Use of extremely short F{\"{o}}rster resonance energy transfer probes in real-time polymerase chain reaction.
\textit{Nucleic Acids Res.},
\textbf{41},
e191.

\bibitem{Ruijter2009}
Ruijter,J.M., Ramakers,C., Hoogaars,W.M., Karlen,Y., Bakker,O., van den Hoff,M.J., and Moorman,A.F. (2009)
Amplification efficiency: linking baseline and bias in the analysis of quantitative PCR data.
\textit{Nucleic Acids Res.},
\textbf{37},
e45.

\bibitem{Spiess2016}
Spiess,A., R{\"{o}}diger,S., Burdukiewicz,M., Volksdorf,T., and Tellinghuisen,J. (2016)
System-specific periodicity in quantitative real-time polymerase chain reaction data questions threshold-based quantitation.
\textit{Sci. Rep.},
\textbf{6},
38951.

\bibitem{Luu-The2005}
Luu-The,V., Paquet,N., Calvo,E., and Cumps,J. (2005)
Improved real-time RT-PCR method for high-throughput measurements using second derivative calculation and double correction.
\textit{BioTechniques},
\textbf{38},
287--293.

\bibitem{Rutledge2008}
Rutledge,R.G. and Stewart,D. (2008)
Critical evaluation of methods used to determine amplification efficiency refutes the exponential character of real-time PCR.
\textit{BMC Mol. Biol.},
\textbf{9},
96.

\bibitem{Gevertz2005}
Gevertz,J.L., Dunn,S.M., and Roth,C.M. (2005)
Mathematical model of real-time PCR kinetics.
\textit{Biotechnol. Bioeng.},
\textbf{92},
346--355.

\bibitem{Pfaffl2001}
Pfaffl,M.W. (2001)
A new mathematical model for relative quantification in real-time RT-PCR.
\textit{Nucleic Acids Res.},
\textbf{29},
e45.

\bibitem{Thellin1999}
Eisenberg,E. and Levanon,E.Y. (2013) 
Human housekeeping genes, revisited. 
\textit{Trends Genet.}, 
\textbf{29}, 
569--74.

\bibitem{Rutledge2004}
Rutledge,R.G. (2004)
Sigmoidal curve-fitting redefines quantitative real-time PCR with the prospective of developing automated high-throughput applications.
\textit{Nucleic Acids Res.},
\textbf{32},
e178.

\bibitem{Guescini2008}
Guescini,M., Sisti,D., Rocchi,M.BL, Stocchi,L., and Stocchi,V. (2008)
A new real-time PCR method to overcome significant quantitative inaccuracy due to slight amplification inhibition.
\textit{BMC Bioinformatics},
\textbf{9},
326.

\bibitem{Lievens2011}
Lievens,A., Aelst,S.V., den Bulcke,M.V. and Goetghebeur,E. (2011)
Enhanced analysis of real-time PCR data by using a variable efficiency model: FPK-PCR.
\textit{Nucleic Acids Res.},
\textbf{40},
e10.

\bibitem{Goll2006}
Goll,R., Olsen,T., Cui,G. and Florholmen,J. (2006)
Evaluation of absolute quantitation by nonlinear regression in probe-based real-time PCR.
\textit{BMC Bioinformatics},
\textbf{7},
107.

\bibitem{Bar2011}
Bar,T., Kubista,M., and Tichopad,A. (2011)
Validation of kinetics similarity in qPCR.
\textit{Nucleic Acids Res.},
\textbf{40},
1395--1406.

\bibitem{Cobbs2012}
Cobbs,G. (2012)
Stepwise kinetic equilibrium models of quantitative polymerase chain reaction.
\textit{BMC Bioinformatics},
\textbf{13},
203.

\bibitem{Stone2006}
Stone,E., Goldes,J. and Garlick,M. (2006)
A two stage model for quantitative PCR.
\textit{The University of Montana, Department of Mathematical Sciences, technical report}.

\bibitem{Lalam2008}
Lalam,N. (2008)
A quantitative approach for polymerase chain reactions based on a hidden Markov model.
\textit{J. Math. Biol.},
\textbf{59},
517--533.

\bibitem{Sivaganesan2008}
Sivaganesan,M., Seifring,S., Varma,M., Haugland,R.A. and Shanks,O.C. (2008)
A Bayesian method for calculating real-time quantitative PCR calibration curves using absolute plasmid DNA standards.
\textit{BMC Bioinformatics},
\textbf{9},
120.

\bibitem{Carr2012}
Carr,A.C. and Moore,S.D. (2012)
Robust Quantification of Polymerase Chain Reactions Using Global Fitting.
\textit{PLoS One},
\textbf{7},
e37640.

\bibitem{Marimuthu2014}
Marimuthu,K., Jing,C. and Chakrabarti,R. (2014)
Sequence-dependent biophysical modeling of DNA amplification.
\textit{Biophys. J.},
\textbf{107},
1731--1743.

\end{thebibliography}







\end{document}